\documentclass[a4paper, aps, prc, showpacs, superscriptaddress, preprintnumbers, nofootinbib, twocolumn]{revtex4-2}
\usepackage[pdftex]{graphicx, color}
\usepackage{graphbox, amsmath, amssymb, amscd, latexsym, bm, braket}
\usepackage[colorlinks=true,pdfstartview=FitV,linkcolor=blue,citecolor=magenta,urlcolor=blue,bookmarks=true,bookmarksnumbered=true]{hyperref}

\begin{document}
\title{Estimation of electric field in intermediate-energy heavy-ion collisions}

\author{Hidetoshi Taya}
\email{hidetoshi.taya@riken.jp}
\address{RIKEN iTHEMS, RIKEN, Wako 351-0198, Japan}

\author{Toru Nishimura}
\address{Department of Physics, Osaka University, Toyonaka, Osaka, 560-0043, Japan}
\address{Yukawa Institute for Theoretical Physics, Kyoto University, Kyoto, 606-8317, Japan}

\author{Akira Ohnishi}
\thanks{Deceased.  }
\address{Yukawa Institute for Theoretical Physics, Kyoto University, Kyoto, 606-8317, Japan}

\begin{abstract}
We estimate the spacetime profile of the electric field in head-on heavy-ion collisions at intermediate collision energies $\sqrt{s_{\rm NN}} = {\mathcal O}(3 \;\mathchar`-\; 10\;{\rm GeV})$.  Using a hadronic cascade model (JAM; Jet AA Microscopic transport model), we numerically demonstrate that the produced field has strength $eE = {\mathcal O}((30 \;\mathchar`-\; 60\;{\rm MeV})^2)$, which is supercritical to the Schwinger limit of QED and is non-negligibly large compared even to the hadron/QCD scale, and survives for a long time $\tau = {\mathcal O}(10\;{\rm fm}/c)$ due to the baryon stopping.  We show that the produced field is nonperturbatively strong in the sense that the nonperturbativity parameters (e.g., the Keldysh parameter) are sufficiently large.  This is in contrast to high-energy collisions $\sqrt{s_{\rm NN}} \gtrsim 100 \;{\rm GeV}$, where the field is extremely short-lived and hence is perturbative.  Our results imply that the electromagnetic field may have phenomenological impacts on hadronic/QCD processes in intermediate-energy heavy-ion collisions and that heavy-ion collisions can be used as a new tool to explore strong-field physics in the nonperturbative regime.  
\end{abstract}

\maketitle

\section{Introduction} \label{sec:1}

Super dense matter, such as that realized inside a neutron star or even denser, can be produced on Earth by colliding heavy ions at intermediate collision energies $\sqrt{s_{\rm NN}} = {\mathcal O}(3\;\mathchar`-\;10 \; {\rm GeV})$.  Such collision experiments have been performed in the Beam Energy Scan program at RHIC~\cite{Aparin:2023fml} and are planned worldwide (e.g., FAIR~\cite{FAIR}, NICA~\cite{NICA}, HIAF~\cite{HIAF}, J-PARC-HI~\cite{J-Parc-HI}) to reveal the extreme form of matter in the dense limit and to develop a better understanding of strong interaction, or quantum chromodynamics (QCD).  These experimental programs have motivated various theoretical studies, which are mainly aimed at investigating the consequences of the high-density matter and the dynamics of how it can be created during the collisions, e.g., novel phases of QCD at finite density (see Ref.~\cite{Fukushima:2010bq} for a review) and the development of various transport models to simulate the realtime collision dynamics such as RQMD~\cite{Sorge:1995dp}, UrQMD~\cite{Bass:1998ca, Bleicher:1999xi}, JAM~\cite{Nara:1999dz}, and SMASH~\cite{SMASH:2016zqf}.  

The purpose of this paper is, rather than pursuing the high-density physics as previously discussed, to point out that a strong electromagnetic field can be created in intermediate-energy heavy-ion collisions.  The generation of such a strong electromagnetic field is of interest not only to hadron/QCD physics but also to the area of strong-field physics.  For hadron/QCD physics, electromagnetic observables such as di-lepton yields~\cite{Sasaki:2019jyh, Savchuk:2022aev, Nishimura:2023oqn, Nishimura:2023not}, which are promising probes of nontrivial processes induced by the high-density matter, are naturally affected by the presence of a strong electromagnetic field.  A correct estimation of the electromagnetic-field profile (and also its implementation into transport-model simulations; cf. Ref.~\cite{Ou:2011fm, Sun:2019hao, Wei:2021yiy}, with emphasis on magnetic-field effects) is, therefore, important when extracting/interpreting signals of the high-density matter from the actual experimental data.  In addition, a strong electromagnetic field can also lead to nontrivial dynamics in the collisions such as the charged flow~\cite{Gursoy:2014aka, Gursoy:2018yai} and therefore is of interest in its own right (see Refs.~\cite{STAR:2023jdd, Dash:2024qcc} for the recent experimental confirmation in the STAR experiment at RHIC).  As for strong-field physics, the generation of a strong electromagnetic field would provide a unique and novel opportunity to study quantum electrodynamics (QED) in the nonperturbative regime beyond the Schwinger limit $eE_{\rm cr} := m_e^2 = (0.511\;{\rm MeV})^2 $ (with $e = |e|$ being the elementary electric charge, $E$ electric field strength, and $m_e$ the electron mass).  Currently, strong-field physics is driven mainly by high-power lasers (see, e.g., Refs.~\cite{DiPiazza:2011tq, Fedotov:2022ely} for reviews).  The focused laser intensity of $I = 1 \times 10^{23}\;{\rm W/cm^2}$ (corresponding to $E \approx 10^{-3}\,E_{\rm cr} $) is the current world record~\cite{Yoon:21}, which is envisaged to be surpassed by the latest and future facilities such as Extreme Light Infrastructure (ELI) with $I = {\mathcal O}(10^{25}\;{\rm W/cm^2})$~\cite{ELI}.  Although the laser intensity is growing rapidly, it is and will remain, for at least the next decade, several orders of magnitude below the Schwinger limit $E_{\rm cr}$.  Therefore, it is difficult to study strong-field phenomena with current lasers.  This means that a novel method or physical system to realize a strong electromagnetic field is highly demanded. 

There exist a number of studies on the generation of a strong electromagnetic field at low- and high-energies both theoretically and experimentally.  Let us briefly review them, so as to clarify our motivation to go to the intermediate energy.  At low energies, due to the baryon stopping (i.e., the Landau picture~\cite{10.1143/ptp/5.4.570, Landau:1953gs}), the collided ions stick together at the collision point and form up a gigantic ion with large atomic number $Z = {\mathcal O}(100)$, and thereby creates a strong Coulomb electric field of the order of $eE \sim (e^2/4\pi) Z/R^2 = {\mathcal O}((20\;{\rm MeV})^2)$, where $R = {\mathcal O}(10\;{\rm fm})$ is the typical radius of the gigantic ion.  The produced field is weak compared to the hadron/QCD scale but is far surpassing the Schwinger limit of QED.  Thus, it is expected to induce intriguing nonlinear QED processes such as the vacuum decay (see Ref.~\cite{Maltsev:2019ytv} for a recent analysis), experimental investigation of which has been done around 1980s but is not conclusive yet (see, e.g., Ref.~\cite{Rafelski:2016ixr} for possible interpretations of the experimental results).  On the other hand, at high energies, the Bjorken picture~\cite{Bjorken:1982qr} is valid rather than the Landau picture.  The colliding ions penetrate with each other without sticking, and hence they do not form up a gigantic ion with large $Z$, unlike the low-energy case.  Nevertheless, high-energy heavy-ion collisions are able to produce a very strong electromagnetic field with a different mechanism.  Namely, due to a strong Lorentz contraction, the charge density $\rho$ of the incident ions are enhanced by the Lorentz factor $\gamma \approx (\sqrt{s_{\rm NN}}/1\;{\rm GeV})/2$ as $\rho \to \gamma \rho \approx \gamma e\rho_0/2$, where $\rho_0 \approx 0.168\;{\rm fm}^{-3}$ is the nuclear saturation density and we halved it because roughly the half of the ions is composed of charged nucleons (i.e., proton).  Due to this enhancement, at central collision events, a strong Coulomb electric field with peak strength $eE \sim e\rho R$, which can exceed the hadron/QCD scale at the RHIC/LHC energy scale $\sqrt{s_{\rm NN}} = {\mathcal O}(100\;\mathchar`-\;1000 \; {\rm GeV})$ and can achieve $eE = {\mathcal O}(100\;\mathchar`-\;1000 \; {\rm MeV})$~\cite{Deng:2012pc}, is produced at the instant of when the colliding ions maximally overlap with each other.  Note that, over the past decade, there have been intensive studies on the generation of a strong {\it magnetic} field in non-central high-energy heavy-ion collisions (see Ref.~\cite{Hattori:2016emy} as a review and references therein), in connection to the search of the chiral magnetic effect~\cite{Kharzeev:2007jp, Fukushima:2008xe}.  Such a strong magnetic field is produced via the Amp\`{e}re law (and subsequently induce a strong electric field due to the time-dependence of the magnetic field through the Faraday law~\cite{Voronyuk:2011jd, Bzdak:2011yy, Deng:2012pc}) and is experimentally utilized to test intriguing nonlinear QED processes such as the photon-photon scattering~\cite{ATLAS:2017fur} and the linear Breit-Wheeler process~\cite{STAR:2019wlg, Brandenburg:2022tna}.  We emphasize that the scope of the present paper is very different to those magnetic studies.  What we are interested in is the {\it electric} field, which is produced by the Coulomb law and is maximized at central events and where the collision energy is not high such that the baryon stopping is effective.

Besides the field strength, there is a crucial difference between the low- and high-energy cases: the lifetime of the produced field.  The lifetime of the field is long at low energies due to the baryon stopping.  For example, it has been shown that the lifetime $\tau$ can reach $\tau = {\mathcal O}(100\;\mathchar`-\;1000\;{\rm fm}/c)$ for collision energies close to the Coulomb barrier~\cite{Maruyama:2001tb, Zagrebaev:2006wp, Tian:2008zza}.  In contrast, the lifetime is extremely short at high energies.  The produced field can survive only for the instance that the colliding ions pass through each other, which is strongly suppressed by the Lorentz factor as $\tau \sim R/\gamma = {\mathcal O}(0.1\;\mathchar`-\;0.01\;{\rm fm}/c)$ at the RHIC/LHC energy.  

The shortness of the lifetime $\tau$ significantly affects the {\it nonperturbativity} of the physics induced by the strong field.  Namely, no matter how strong a field is, the physics has to be perturbative if it is short-lived.  The low-order perturbation theory becomes sufficient in such a limit, and therefore the physics becomes ``trivial'' in the sense that it is not very different from the usual electromagnetic processes in the vacuum.  Intuitively, this is simply because there is no time for the finite field to have multiple interactions with a particle.  Such an idea has been developed mainly in the context of laser or strong-field QED, but let us here introduce it to the context of heavy-ion collisions.  To be more quantitative, suppose we have, as an example, an electric field with peak strength $E_0$ and lifetime $\tau$, and consider the vacuum pair production by such a strong electric field, ${\rm vacuum} + E \to e^+ + e^-$.  For this case, the interplay between the nonperturbative and perturbative pair production is controlled by two dimensionless quantities~\cite{Popov:1971, Popov:1971iga, 1970PhRvD...2.1191B, Dunne:2005sx, Dunne:2006st, Oka:2011ct, Taya:2014taa, Gelis:2015kya, Aleksandrov:2018zso, Taya:2020dco}, 
\begin{align}
	\xi(m) := \frac{eE_0\tau}{m} \ \ {\rm and}\ \ 
	\nu := eE_0 \tau^2 \;. \label{eq:01}
\end{align}
Note that the parameter $\xi$ is known as the Keldysh parameter~\cite{Keldysh:1965ojf} and, depending on the context, is also called the classical nonlinearity parameter~\cite{Fedotov:2022ely}.  The mass $m$ is the mass of the particle to be produced.  For enough strong and long-lived fields such that $\xi, \nu \gg 1$, the pair production becomes nonperturbative in the sense that the rate of the pair production acquires a non-analytic dependence in $e$ and $E_0$ as $\propto \exp[-({\rm const.}) \times (m^2/eE_0)]$.  In the opposite limit, $\xi, \nu \ll 1$, it becomes purely perturbative, i.e., the low-order perturbative treatment becomes sufficient, meaning that the rate only has the power dependence in $e$ and $E_0$ as $\propto (eE_0/m^2)^n$ ($n \in {\mathbb N}$).  This example of the vacuum pair production clearly demonstrates that the supercriticality $eE_0 \gtrsim m^2$ is not sufficient to guarantee the nonperturbativity of strong-field processes.  One must, thus, pay attention to the magnitude of the lifetime $\tau$, or the resulting $\xi$ and $\nu$, as well, in addition to the strength $E_0$.  In terms of these nonperturbativity parameters (\ref{eq:01}), high-energy heavy-ion collisions at the typical RHIC/LHC scale correspond to $\nu = {\mathcal O}(0.1)$ and $\xi = {\mathcal O}(100)$ for electron $m=m_e$ and ${\mathcal O}(1)$ for pion $m = m_\pi \approx 140\;{\rm MeV}$.  This means, although $\xi$ can be nonperturbatively large for the QED scale $m=m_e$ due to the largeness of the field strength, $\nu$ always remains perturbatively small due to the shortness of the lifetime.  For the hadron/QCD scale $m=m_\pi$, neither $\xi$ nor $\nu$ can be nonperturbative.  Therefore, in either case of QED or QCD, the pair production cannot be nonperturbative, implying that strong-field physics in the nonperturbative regime cannot be explored with the field produced in high-energy heavy-ion collisions.  This naive argument, based on the ``order parameters'' of the vacuum pair production, is consistent with the actual experimental results.  There have been observations of next-to-leading-order QED processes such as light-by-light scattering~\cite{ATLAS:2017fur} and linear Breit-Wheeler pair production~\cite{STAR:2019wlg}, in which the experimental results are consistent with perturbative QED calculations and no signatures of higher-order nonlinear effects have been detected.  

We are now in a position to address the advantage of going to intermediate energies.  It is natural to expect that the electromagnetic field generated in intermediate-energy heavy-ion collisions should have characteristics between the low- and high-energy cases.  Namely, although the field strength would be weaker than that in the high-energy case, it should be stronger than that in the low-energy case.  This means that the produced field will remain much stronger than the Schwinger limit of QED and may still be comparable to the hadron/QCD scale.  As for lifetime, we can naturally expect that the field produced at intermediate energies should survive longer than the high-energy field.  If this is true, it overcomes the problem of the short lifetime in the high energy limit, while maintaining a sufficiently large field strength.  Hence, it enables us to access the strong-field physics in the nonperturbative regime.  It is therefore worthwhile to investigate the generation of a strong electromagnetic field in intermediate-energy heavy-ion collisions and to make a realistic estimate to clarify whether this is the case, and, if so, how important it is.  To the best of our knowledge, there does not exist such a quantitative estimation of the electromagnetic field at intermediate energies.  

Based on the motivations explained thus far, we would like to make a realistic estimate of the electromagnetic field in intermediate-energy heavy-ion collisions based on a hadron-transport-model simulation.  Indeed, neither the Landau nor the Bjorken picture is complete at intermediate energies (cf. Ref.~\cite{Deng:2020ygd} as a related for vorticity estimation).  Accordingly, the generation mechanism of the electromagnetic field is more involved compared to the low- and high-energy limits, and hence it is impossible to carry out a simple analytic estimate at intermediate energies.  In particular, dynamical effects become more important than the other energy regimes.  For example, both elastic- and inelastic- multiple collision processes among the constituent nucleons are important to determine the sticking or non-sticking of the colliding ions.  Hadrons are produced or decay during the time evolution through the inelastic processes, meaning that the simple but widely-used modeling based on the Li\'{e}nard-Wiechert potential of the electromagnetism is not applicable\footnote{The Li\'{e}nard-Wiechert potential is explicitly dependent on time-derivatives of the trajectory of a charged particle ${\bm r}(t)$.  Therefore, it is valid only if ${\bm r}(t)$ does not have any singularities, which means the particle cannot decay nor be produced during the time evolution.  }.  The Lorentz contraction comes into play, unlike at the low energy limit, but is not as strong as that at the high energy, and therefore hadrons consisting of the colliding ions need to be treated as anisotropic objects with finite size.  All of those dynamical effects can be taken into account systematically by using established hadron-transport models of heavy-ion collisions.  In this work, we adopt JAM (Jet AA Microscopic transport model)~\cite{Nara:1999dz, JAM}.  JAM is a hadronic cascade model to simulate the realtime dynamics of heavy-ion collisions, which is applicable to a wide range of collision energies from $\sqrt{s_{\rm NN}} = {\mathcal O}(2\;{\rm GeV})$ up to ${\mathcal O}(100\;{\rm GeV})$.  More details of JAM, including its comparison with other models, can be found in, e.g., Ref.~\cite{TMEP:2022xjg}, and also in the main text below.  

This paper is organized as follows.  We first explain our numerical setup and how we calculate the electromagnetic field with JAM in Sec.~\ref{sec:2}.  We then present the numerical results in Sec.~\ref{sec:3} that include the spacetime profiles of the charge density (Sec.~\ref{sec:3A}) and the resulting electromagnetic field (Sec.~\ref{sec:3B}), the time evolution of the peak strength (Sec.~\ref{sec:3C}), and the corresponding nonperturbativity parameters $\xi$ and $\nu$ to demonstrate that the field is nonperturbatively strong (Sec.~\ref{sec:3D}).

{\it Notation}: Our metric is the mostly minus, i.e., $g^{\mu\nu} := {\rm diag}(+1,-1,-1,-1)$.  Spatial three vectors are indicated by the bold letters, e.g., $x^\mu := (t, {\bm x})$ for spacetime coordinates and $A^\mu = (A^0, {\bm A})$ for the four-vector potential.  We take the $z$-axis along the beam direction and write ${\bm x} =: (x,y,z) =: ({\bm x}_\perp,z)$.

\section{Numerical recipe} \label{sec:2}

\subsection{Calculation program} \label{sec:2a}

We wish to calculate the electric ${\bm E}$ and magnetic ${\bm B}$ fields in heavy-ion collisions at intermediate collision energies.  The electromagnetic fields are given by 
\begin{align}
	{\bm E} := -\partial_t {\bm A} - {\bm \nabla} A^0 \ \ {\rm and}\ \ 
	{\bm B} := {\bm \nabla} \times {\bm A} \;, \label{eq:1}
\end{align}
where $A^\mu$ is the retarded vector potential,
\begin{align}
	A^\mu (t,{\bm x}) := \frac{1}{4\pi} \int {\rm d}^3{\bm x}' \frac{J^\mu(t-|{\bm x}-{\bm x}'|,{\bm x}')}{|{\bm x}-{\bm x}'|} \;,  \label{eq:2}
\end{align}
with $J^\mu$ being the total electric current carried by the charged particles in the collisions.  In the present paper, we focus on the event-averaged values of $O = {\bm E},{\bm B}$ and take the event averaging of Eq.~(\ref{eq:1}) as
\begin{align}
	O := \frac{1}{N} \sum_{i=1}^N O_i \;, \label{eq:3}
\end{align}
where $O_i$ is the event-by-event result and $N$ is the total number of the events.

We evaluate the electromagnetic field (\ref{eq:1}) numerically as follow.  We first calculate the electric current $J^\mu$ with JAM.  JAM calculates the phase-space distributions of hadrons in heavy-ion collisions $({\bm x}_n(t), p^\mu_n(t))$, where the subscript $n$ labels the hadrons in a collision event.  The electric current $J^\mu$ at a position $(t',{\bm x}')$ is then given in terms of the JAM phase-space distribution as
\begin{align}
	J^\mu(t',{\bm x}') := \sum_{n \in {\rm all\;hadrons}} \frac{p^\mu_n(t')}{p^0_n(t')} \rho_n(t',{\bm x}') \;, \label{eq:4}
\end{align}
where $\rho_n$ is the charge density of a single charged hadron.  In JAM, hadrons are treated as if they are point-like, and hence the charge density of each hadron is localized strictly at ${\bm x}_n$.  Hadrons are finite-sized in reality, motivated by which we smear the charge density by using the (relativistic) Gaussian distribution:
\begin{align}
	&\rho_n(t',{\bm x}')
	:= q_n \frac{\gamma_n(t')}{(\sqrt{2\pi}\sigma)^3} \nonumber\\
	&\times \exp\left[ - \frac{ |{\bm x}'-{\bm x}_n(t')|^2 + \gamma_n^2(t') \left( {\bm v}_n(t')\cdot ({\bm x}'-{\bm x}_n(t')) \right)^2}{2\sigma^2}  \right] \;, \nonumber 
\end{align}
where $q_n$ is the electric charge, $\sigma$ the smearing width, and $\gamma := \sqrt{1 - {\bm v}_n^2 } := \sqrt{1 - ({\bm p}_n/p^0_n)^2 } $ the Lorentz factor for the $n$-th hadron.

Once the electric current $J^\mu$ is obtained with JAM, can we calculate the retarded vector potential $A^\mu(t,{\bm x})$ by carrying out the ${\bm x}'$ integration according to Eq.~(\ref{eq:2}) at each spacetime point $(t,{\bm x})$.  This can be done with the standard numerical integration schemes, e.g., the Gauss-Legendre quadrature method, which we have adopted in this work.

The simulations start from some initial time $t_{\rm in} =: 0$, which is defined to be the time when the colliding ions get as close as $3.0\; {\rm fm}$ in the default setting of JAM.  The JAM hadron phase-space data are, thus, available only later than $t_{\rm in}$, so is the electric current $J^\mu$ (\ref{eq:4}).  To perform the integration (\ref{eq:2}), the information of the electric current at time $t'=t-|{\bm x}-{\bm x}'|$, which can be earlier than $t_{\rm in}$ (when $|{\bm x}-{\bm x}'| > t-t_{\rm in}$), is needed.  To get the electric current before the initial time $t_{\rm in}$, we extrapolate the phase-space data at the initial time $t=t_{\rm in}$ by assuming that the hadrons that constitutes the incident ions go straight trajectories in the phase-space without any interactions: 
\begin{align}
\begin{split}
	&{\bm x}_n(t<t_{\rm in}) := {\bm x}_n(t_{\rm in}) + (t-t_{\rm in}) \frac{{\bm p}_n(t_{\rm in})}{p^0_n(t_{\rm in})} \;, \\
	&p^\mu_n(t<t_{\rm in}) := p^\mu_n(t_{\rm in}) \;.
\end{split}
\end{align}

Having calculated the retarded vector potential $A^\mu$ (\ref{eq:2}), can we obtain the electromagnetic fields via the differentiation (\ref{eq:1}).  We in this work have adopted the standard central difference method with discretizing the ${\bm x}$ coordinates with uniform meshes.

\subsection{Numerical parameters} \label{sec:2b}

\begin{table*}[!t]
\begin{center}
\begin{tabular}{lcl} 
	\hline\hline
	&&\\[-3mm]
	Ion	 species										& & ${}^{197}{\rm Au}$ \\
	Collision energy $\sqrt{s_{\rm NN}}$		& & 2.4, 3.0, 3.5, 3.9, 4.5, 5.2, 6.2, 7.2, 7.7, 9.2, and 11.5\;{\rm GeV} \\ 
	Impact parameter $b$							& & 0\;{\rm fm} \\ 
	Gaussian smearing width $\sigma$			& & 1\;{\rm fm} \\
	Total event number $N$							& & 100 \\
	Discretization of time $t$					& & $t\in[0,30]\;{\rm fm}/c$ with $\Delta t = 0.5\;{\rm fm}/c$ \\
	Discretization of ${\bm x}=(x,y,z)$			& & $x,y,z\in [-30,+30]\;{\rm fm}$ with $(n_x,n_y,n_z) = (121,121,121)$ sites\\&& (a uniform 3D lattice with $\Delta x = \Delta y = \Delta z = 0.5\;{\rm fm}$) \\
	Discretization of ${\bm x}'=(x',y',z')$	& & $x',y',z'\in [-30,+30]\;{\rm fm}$ with $(n_{x'},n_{y'},n_{z'}) = (120,120,120)$ sites\\&& (a non-uniform 3D lattice for the Gauss-Legendre quadrature method) \\
	&&\\[-3mm]
	\hline\hline
\end{tabular}
\caption{Parameters adopted in the numerical simulation.  }\label{table:1}
\end{center}
\end{table*}

Using JAM, we simulate head-on (i.e., the impact parameter $b$ is strictly fixed at $b=0$) collisions of gold ions at the intermediate energies $\sqrt{s_{\rm NN}} = 2.4, 3.0, 3.5, 3.9, 4.5, 5.2, 6.2, 7.2, 7.7, 9.2$, and $11.5$.  The purpose of this paper is to present {\it a baseline estimation} of the electromagnetic field produced in the so-far unexplored intermediate-energy regime.  For this reason, we shall switch off all the advanced options of JAM (such as the mean-field potential and hybrid hydrodynamic simulation) and carry out simulations with the default JAM setting.  In other words, our estimation is based on the basic hadronic cascade model, which has been adopted widely in heavy-ion simulations, and is less sensitive to the modeling details of JAM.  

We summarize the numerical parameters adopted in the present work in Table~\ref{table:1}.  We have five numerical parameters in the evaluation of the electromagnetic field outlined in Sec.~\ref{sec:2a}: the event number $N$, the smearing width $\sigma$, and the discretizations of $t, {\bm x}$, and ${\bm x}'$.  

Several further comments are in order about our numerical parameters and settings: 
\begin{itemize}
\item Although we shall concentrate on gold collisions as a first step to study the electromagnetic-field physics in intermediate-energy heavy-ion collisions, it is an interesting question to consider other ion species.  In particular, deformed ions such as uranium and also asymmetric collisions such as Au\,+\,Cu (see, e.g., Refs.~\cite{Hirono:2012rt, Voronyuk:2014rna, STAR:2016cio} at high energies) would be of interest, in which case the resulting electromagnetic fields should have a preferred direction due to the collision geometry, leading to observable effects such as a charged $v_1$ flow.  

\item The impact parameter $b$ dependence is also an interesting issue, although we shall concentrate on the central events $b=0$ in what follows.  For finite impact parameters, magnetic field shall dominate, while electric field does so in head-on events as we shall demonstrate below.  We shall report the interplay of the dominance between the electric and magnetic fields in a future publication.  

\item Although we shall estimate the electromagnetic fields with the $N=100$ event averaging, event-by-event fluctuations may be important in actual experiments (see also Ref.~\cite{Deng:2012pc} for the high-energy case).  For example, it has been argued in Refs.~\cite{Ohnishi:2002es, ohnishi2016approaches, prepdensity} that the maximum value of the baryon density can fluctuate event-by-event by more than 30\;\% at intermediate energies.  Such large fluctuations of the baryon density $\rho_{\rm B}$ imply that the charge density $\rho/e \approx \rho_{\rm B}/2$ should also fluctuate, and so does the resulting electromagnetic field.  We shall report the event-by-event fluctuations of the electromagnetic fields in intermediate-energy heavy-ion collisions in a forthcoming paper.  Note that we are reluctant to take a relatively small event number $N=100$ in order to reduce the numerical cost, although the residual event-by-event fluctuations after the event averaging do not look significant as we show later.  

\item We shall use the particular value of the smearing width $\sigma = 1\;{\rm fm}$ for all hadron species.  This is physically motivated by the fact that the typical size of a nucleon is of the order of $1\;{\rm fm}$, and this is a common choice in the literature.  We have checked that the numerical results are less sensitive to the value of $\sigma$ after taking the event averaging.  Nonetheless, for a more realistic estimation, in particular for estimating the event-by-event fluctuation, it would be interesting to consider a more realistic value of $\sigma$ as well as to change $\sigma$ depending on hadron species.  

\item Let us comment on some more details of JAM (in its default setting).  JAM is a Monte-Carlo simulation based on the hadoronic cascade model to simulate the realtime dynamics of heavy-ion collisions.  The hadronic cascade model is a model to describe the collision dynamics as a superposition of independent collisions among hadrons, which propagate classically with straight lines in the phase space (which can be curved once the mean-field option, to account for the nontrivial equation of state beyond the ideal gas or quantum effects such as the exchange interaction, is switched on~\cite{Maruyama:1996rn, Isse:2005nk, Mancusi:2009zz, Nara:2019qfd, Nara:2020ztb, Nara:2021fuu}), and do not have the more fundamental quark and gluon degrees of freedom nor do we include the collective hydrodynamic treatment (the latter of which can be included as an advanced option~\cite{Akamatsu:2018olk}).  As the collision channel, not only elastic hadron-hadron collisions but also resonance production, soft-string excitation, and multiple mini-jet production are included as inelastic collisions, which is similar to other hadronic cascade models such as UrQMD~\cite{Bass:1998ca, Bleicher:1999xi}.  Among those inelastic channels, the soft-string excitation, which is implemented according to the Lund model~\cite{Andersson:1997xwk}, dominates in the intermediate-energy regime (in particular for $4\;{\rm GeV} \lesssim \sqrt{s_{\rm NN}} \lesssim 10\;{\rm GeV}$), while, at lower and higher energies, the resonance and the mini-jets become more important, respectively, in JAM.  The collisions occur with the closest distance approach.  It is implemented in JAM in its default setting in a covariant manner~\cite{Nara:2023vrq}, which slightly increases the collision rate compared to the other non-covariant modelings.  

\item Since our estimation is based on the basic hadronic cascade model (i.e., JAM with default setting), various physical effects, especially dense-matter effects that would be important at lower energies (close to $\sqrt{s_{\rm NN}} \approx 2$), are dismissed.  Considering the success of the hadoronic cascade modeling~\cite{TMEP:2022xjg}, we believe that the results that are presented below are sufficient as a baseline.  It is out of scope in the present paper to carry out a more sophisticated transport-model simulation (e.g., with swicthing-on the advanced options of JAM), which is nevertheless important to make a more realistic estimation.  We shall come back to this point and discuss more in Sec.~\ref{sec:4}.  

\item We have carefully checked that the lattice volumes of ${\bm x}$ and ${\bm x}'$ and the mesh sizes $\Delta t, \Delta {\bm x}$, and $\Delta {\bm x}'$ are sufficiently large and fine, respectively, in the sense that the numerical results are not sensitive to them.  For those mesh sizes and the lattice volume, our spacetime is discretized into ${\mathcal O}(100^7)$ meshes, for which a single numerical simulation (per energy) consumes ${\mathcal O}(20\;{\rm GB})$ RAM and takes about a few days with a standard GPU computer available on the market.  

\end{itemize}

\section{Numerical results} \label{sec:3}

We present the numerical results obtained with the recipe outlined in Sec.~\ref{sec:2}.  The main result is that an electric field which is non-perturbatively strong and long-lived can be created in intermediate-energy heavy-ion collisions; see Sec.~\ref{sec:3D}, in particular Fig.~\ref{fig:4}.  Before getting there, we discuss the generation of such a strong electric field step by step.  We first demonstrate in Sec.~\ref{sec:3A} that a highly-charged matter $\rho/e = {\mathcal O}(0.5\;{\rm fm}^{-3})$, as a consequence of the formation of the dense baryon matter, is realized at intermediate energies.  The highly-charged matter is the source of a strong electric field, whose spacetime profile shall be discussed in Sec.~\ref{sec:3B}.  We shall discuss the intensity and the lifetime of the produced electric field in detail in Sec.~\ref{sec:3C} and show in Sec.~\ref{sec:3D} that it can be nonperturbatively strong and long-lived, in contrast to what is realized at high energies.

\subsection{Spacetime profile of the charge density} \label{sec:3A}

\begin{figure*}[!t]
\flushleft{\hspace*{5.2mm}\mbox{$\sqrt{s_{\rm NN}}=2.4\;{\rm GeV}$ \hspace{16.6mm} $3.0\;{\rm GeV}$ \hspace{22.4mm} $5.2\;{\rm GeV}$ \hspace{22.4mm} $6.2\;{\rm GeV}$ \hspace{22.4mm} $7.2\;{\rm GeV}$}} \\
\vspace*{1mm}
\hspace*{-47.7mm}
\includegraphics[align=t, height=0.271\textwidth, clip, trim = 35 0  75 0]{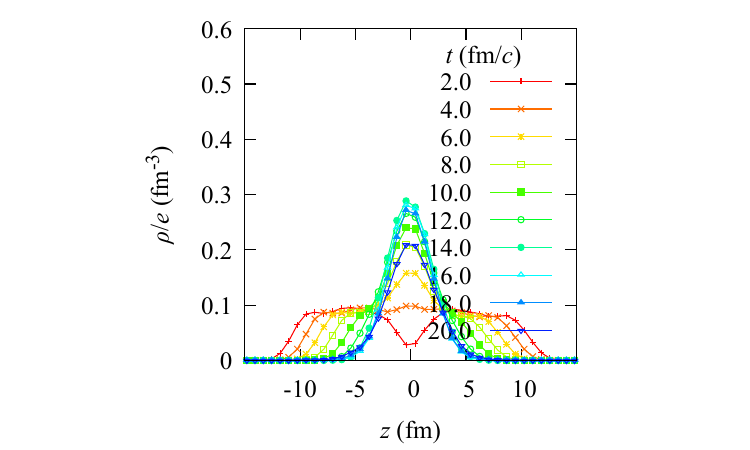}\hspace*{4.3mm}
\includegraphics[align=t, height=0.271\textwidth, clip, trim = 75 0 105 0]{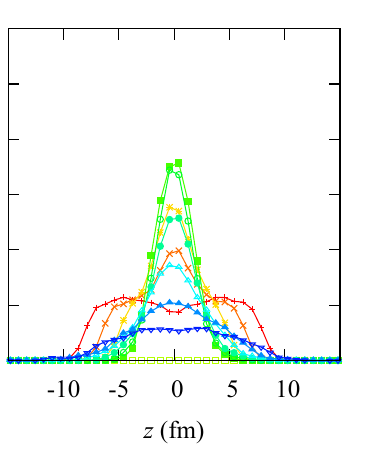}\hspace*{-5.5mm}
\includegraphics[align=t, height=0.271\textwidth, clip, trim = 75 0 105 0]{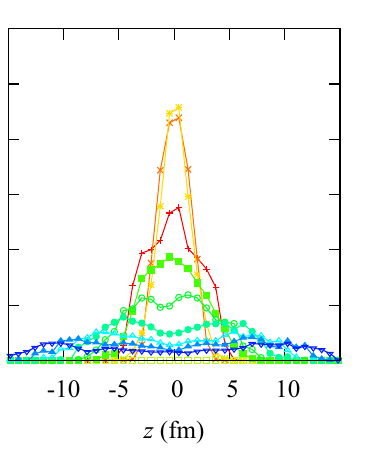}\hspace*{-5.5mm}
\includegraphics[align=t, height=0.271\textwidth, clip, trim = 75 0 105 0]{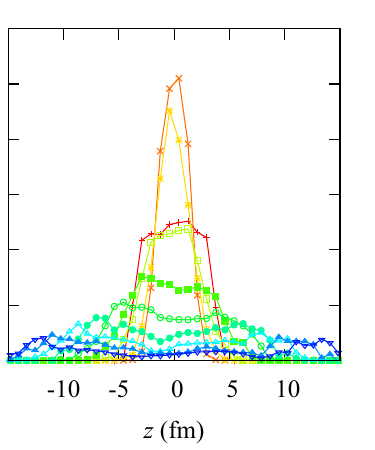}\hspace*{-5.5mm}
\includegraphics[align=t, height=0.271\textwidth, clip, trim = 75 0 105 0]{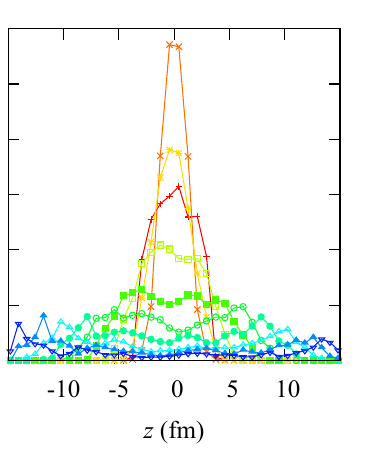}\hspace*{-5.5mm} \\
\hspace*{-47.7mm}
\includegraphics[align=t, height=0.271\textwidth, clip, trim = 35 0  75 0]{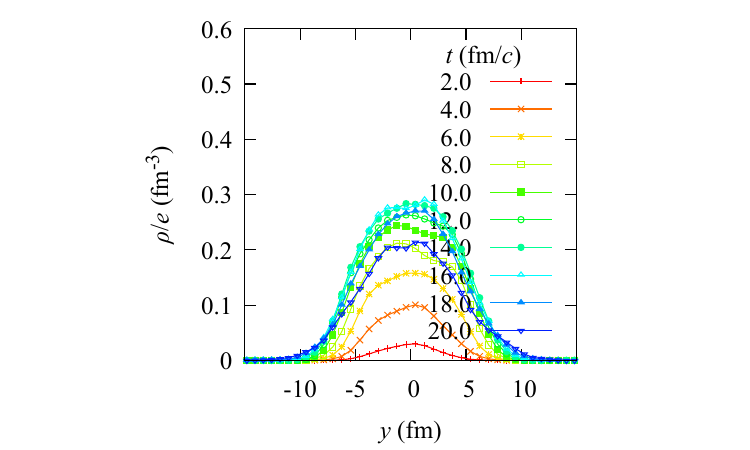}\hspace*{4.3mm}
\includegraphics[align=t, height=0.271\textwidth, clip, trim = 75 0 105 0]{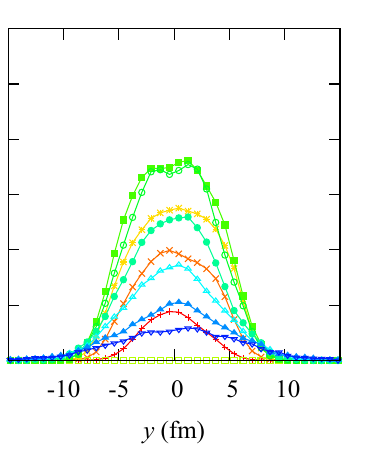}\hspace*{-5.5mm}
\includegraphics[align=t, height=0.271\textwidth, clip, trim = 75 0 105 0]{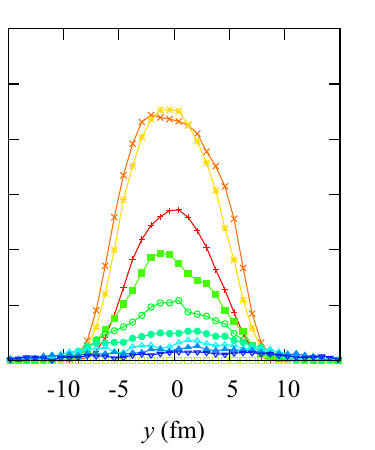}\hspace*{-5.5mm}
\includegraphics[align=t, height=0.271\textwidth, clip, trim = 75 0 105 0]{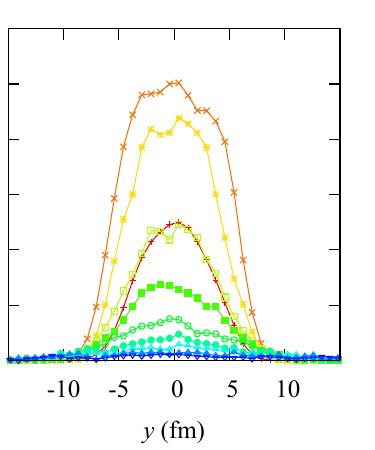}\hspace*{-5.5mm}
\includegraphics[align=t, height=0.271\textwidth, clip, trim = 75 0 105 0]{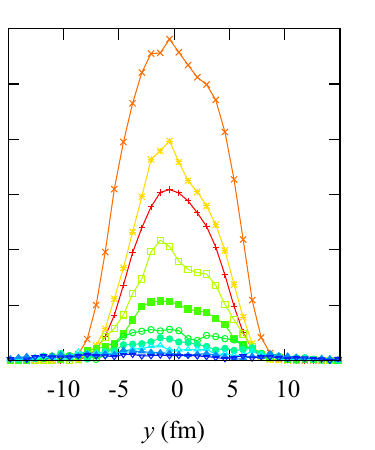}\hspace*{-5.5mm}
\caption{\label{fig:1} The spacetime profile of electric charge density $\rho/e$ at some selected collision energies ($\sqrt{s_{\rm NN}} = 2.4,3.0, 5.2, 6.2$, and $7.2\;{\rm GeV}$).  The different colors indicate different time slices, ranging from $2\;{\rm fm}/c$ (red) to $20\;{\rm fm}/c$ (blue) with the increment of $2\;{\rm fm}/c$.  The top and bottom panels correspond to the distributions in the beam $z$ direction at fixed $x=y=0\;{\rm fm}$ and in the transverse $y$ direction at fixed $x=z=0\;{\rm fm}$, respectively.  
}
\end{figure*}

Figure~\ref{fig:1} shows the spacetime profile of the charge density.  The key observation is that intermediate-energy heavy-ion collisions can realize a high charge density, typically of the order of $\rho/e = {\mathcal O}(0.5\;{\rm fm}^{-3})$, for a relatively long time $\tau = {\mathcal O}(10\;{\rm fm}/c)$.  The magnitude of the charge density is roughly ${\mathcal O}(5)$ times greater than that of an usual charged ion at rest $\rho/e \approx \rho_0/2 \approx 0.08\;{\rm fm}^{-3}$.  This highly-charged matter is the source of a strong electric field, as we shall demonstrate later.    

There are two important physics that contribute to the creation of the highly-charged matter.  The first one is the Lorentz contraction.  Due to the Lorentz contraction in the beam direction, the incident ions are not spherical, but anisotropic, and the longitudinal size is shortened as $R \to R/\gamma \approx (13\;{\rm fm})/(\sqrt{s_{\rm NN}}/1\;{\rm GeV})$, where $R \approx (1.1\;{\rm fm})\times A^{1/3} \approx 6.4\;{\rm fm}$ is the radius of a gold ion $A=197$ at rest and we used $2\gamma \approx (\sqrt{s_{\rm NN}}/1\;{\rm GeV})$.  Accordingly, the charge density is enhanced from that at rest by the Lorentz factor $\gamma$ and hence becomes larger with collision energy as $\rho/e \approx \gamma \times \rho_0/2 \approx (0.04\;{\rm fm}^{-3}) \times (\sqrt{s_{\rm NN}}/1\;{\rm GeV}) $.  The collision dynamics can roughly be understood as an overlapping process of these two Lorentz-contracted ions.  The maximum charge density is then achieved when the two ions maximally overlap with each other.  As a zero-th order approximation, we may neglect the interaction between the ions and hence assume that they just pass through each other during a collision.  Then, the maximum can be estimated as $\max \rho/e \approx 2 \times \rho/e \approx (0.08\;{\rm fm}^{-3}) \times (\sqrt{s_{\rm NN}}/1\;{\rm GeV})$ and the time at which it is achieved as $t_{\rm max} \approx 2R/\gamma \approx (26\;{\rm fm}/c)/(\sqrt{s_{\rm NN}}/1\;{\rm GeV})$.  These estimates are in rough agreement with the numbers shown Fig.~\ref{fig:1}.  

The second is the interaction (i.e., the baryon stopping), which is especially important in determining the lifetime of the highly-charged matter.  The importance of the interaction is evident in Fig.~\ref{fig:1}.  If it is non-interacting, the charge density should exhibit a two-peak structure after the maximum, i.e., the collided ions pass through each other and leave nothing in the mid-rapidity case, $z \approx 0$ (the Bjorken picture~\cite{Bjorken:1982qr}).  This is not the case in Fig.~\ref{fig:1}, which shows that the density is single-peaked at the mid-rapidity after the collision.  In other words, the interaction makes it difficult for the colliding ions to penetrate each other, instead causing them to merge at the point of collision (the Landau picture~\cite{10.1143/ptp/5.4.570, Landau:1953gs}).  The interaction makes the lifetime $\tau$ of the highly-charged matter longer, compared to what is naively expected from the Bjorken picture $\tau \approx 2R/\gamma \approx (26\;{\rm fm}/c)/(\sqrt{s_{\rm NN}}/1\;{\rm GeV})$.  For example, at $\sqrt{s_{\rm NN}} = 3.0\;{\rm GeV}$, a relatively dense charge state $\rho/e \gtrsim 0.1\;{\rm fm}^{-3}$ appears around $t \approx 2\;{\rm fm}/c$ and survives until $t \approx 18\;{\rm fm}/c$, whose lifetime $\tau \approx 18 - 2 \;{\rm fm}/c = 16\;{\rm fm}/c$ is much longer than the naive Bjorken estimate $\tau \approx 26/3\;{\rm fm}/c \approx 9\;{\rm fm}/c$.  The interaction effect is non-negligible and hence a highly-charged matter with a relatively long lifetime is realized in the intermediate energy values $\sqrt{s_{\rm NN}} = {\mathcal O}(3 \;\mathchar`-\; 10\;{\rm GeV})$.

\subsection{Spacetime profile of the produced field} \label{sec:3B}

\begin{figure*}[!t]
\flushleft{\hspace*{6.8mm}\mbox{$\sqrt{s_{\rm NN}}=2.4\;{\rm GeV}$ \hspace{16.6mm} $3.0\;{\rm GeV}$ \hspace{22.1mm} $5.2\;{\rm GeV}$ \hspace{22.1mm} $6.2\;{\rm GeV}$ \hspace{22.1mm} $7.2\;{\rm GeV}$}} \\
\vspace*{1mm}
\hspace*{-47mm}
\includegraphics[align=t, height=0.268\textwidth, clip, trim = 35 0 70 0]{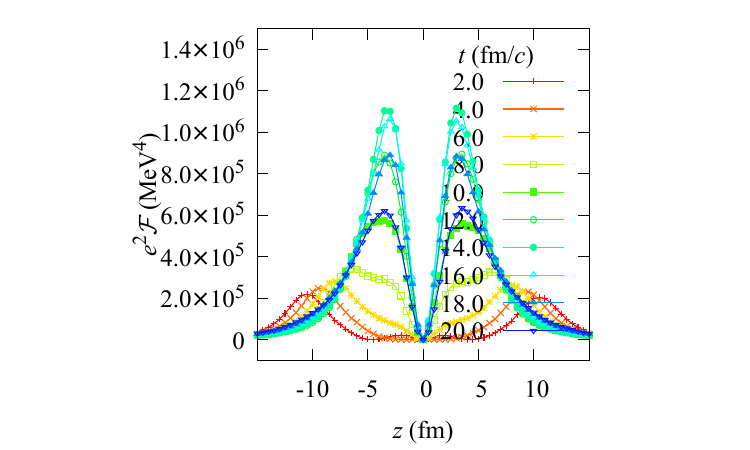}\hspace*{1.5mm}
\includegraphics[align=t, height=0.268\textwidth, clip, trim = 80 0 70 0]{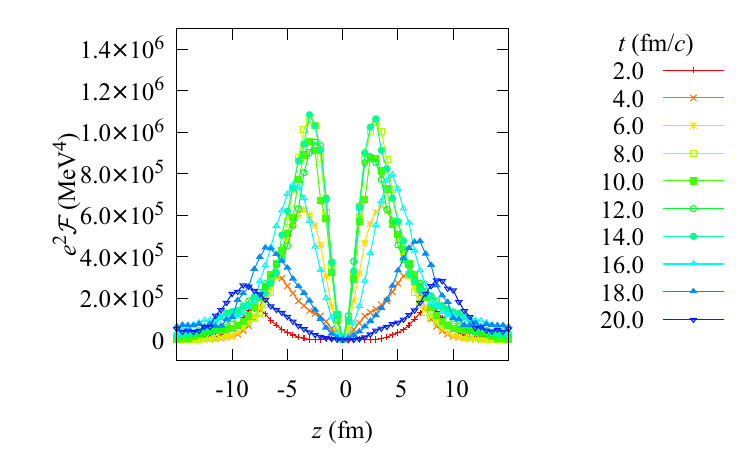}\hspace*{-12.15mm}
\includegraphics[align=t, height=0.268\textwidth, clip, trim = 80 0 70 0]{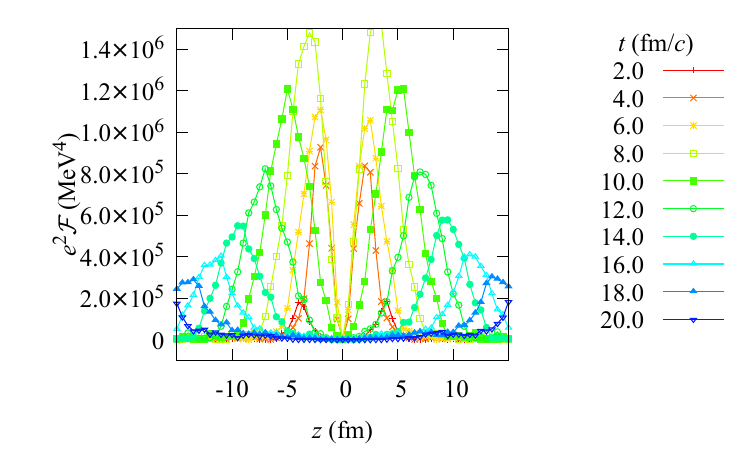}\hspace*{-12.15mm} 
\includegraphics[align=t, height=0.268\textwidth, clip, trim = 80 0 70 0]{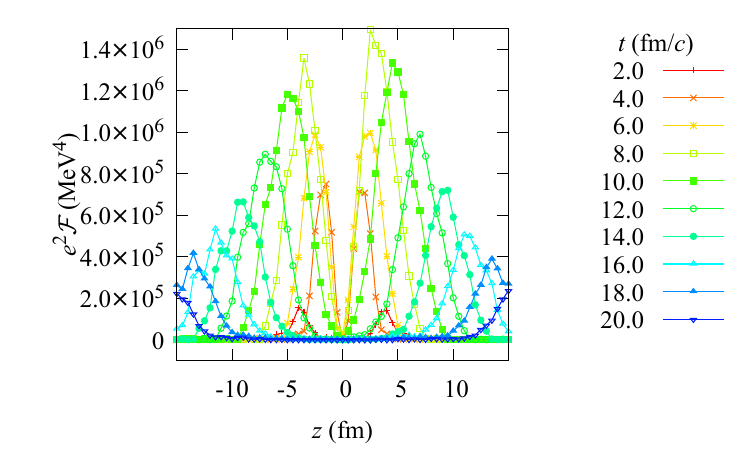}\hspace*{-12.15mm} 
\includegraphics[align=t, height=0.268\textwidth, clip, trim = 80 0 70 0]{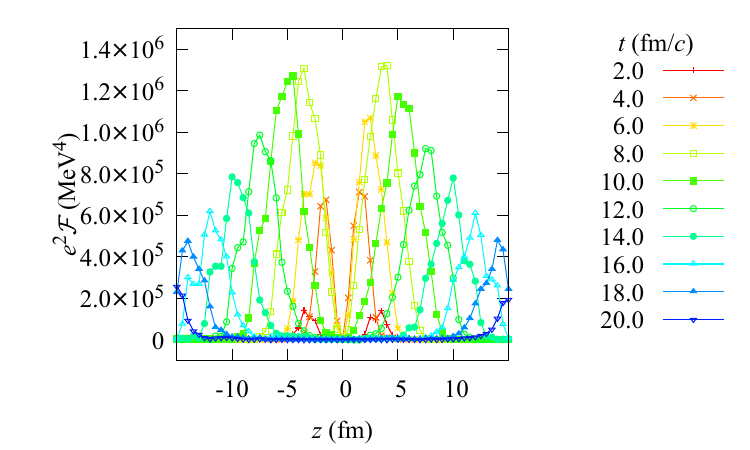}\hspace*{-12.15mm} \\
\hspace*{-47mm}
\includegraphics[align=t, height=0.268\textwidth, clip, trim = 35 0 70 0]{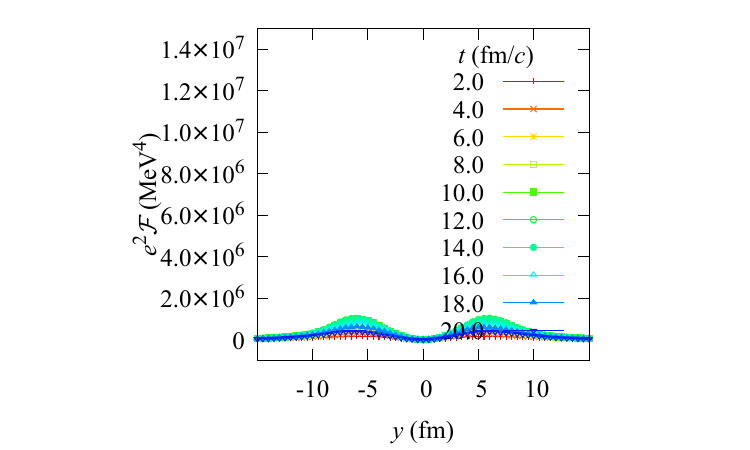}\hspace*{1.5mm}
\includegraphics[align=t, height=0.268\textwidth, clip, trim = 80 0 70 0]{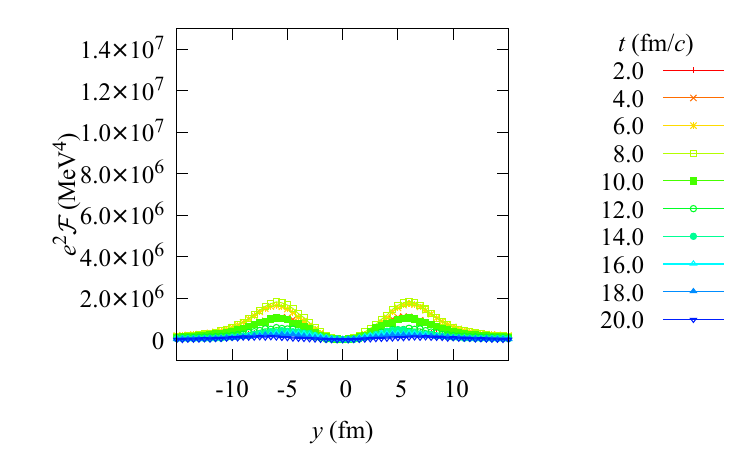}\hspace*{-12.15mm} 
\includegraphics[align=t, height=0.268\textwidth, clip, trim = 80 0 70 0]{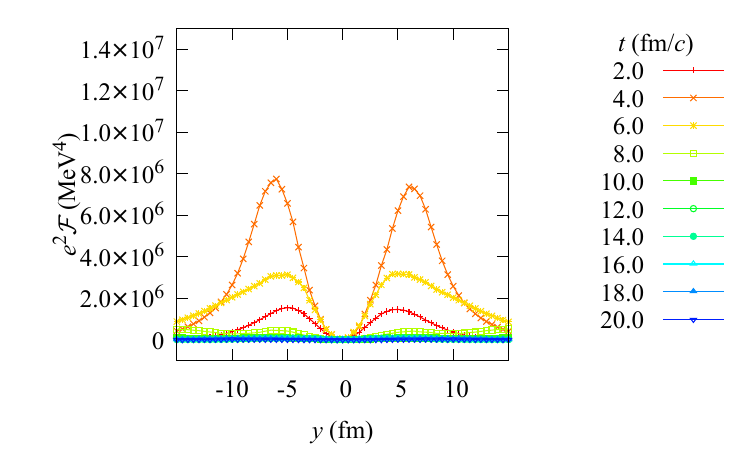}\hspace*{-12.15mm} 
\includegraphics[align=t, height=0.268\textwidth, clip, trim = 80 0 70 0]{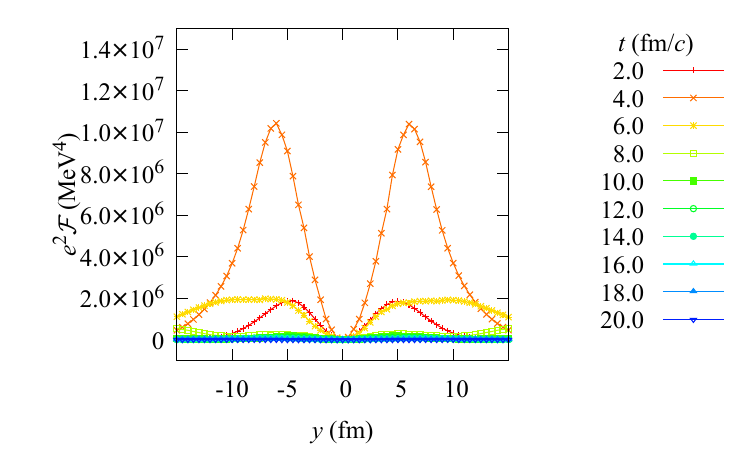}\hspace*{-12.15mm} 
\includegraphics[align=t, height=0.268\textwidth, clip, trim = 80 0 70 0]{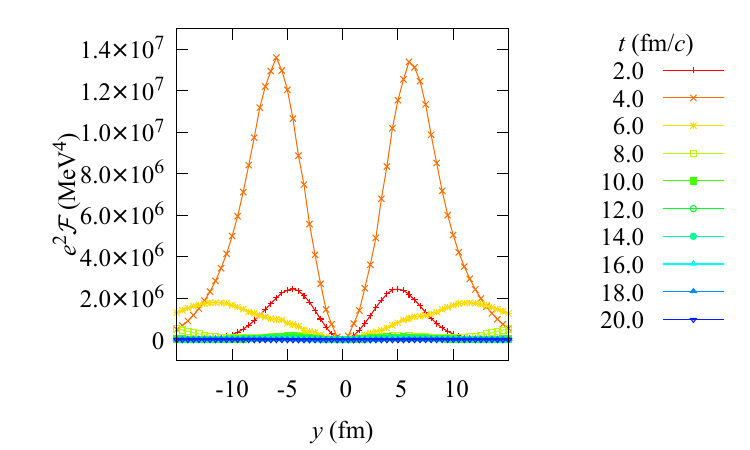}\hspace*{-12.15mm} 
\caption{\label{fig:2} The typical spacetime profile of the Lorentz invariant ${\mathcal F} = {\bm E}^2 - {\bm B}^2$ in intermediate-energy heavy-ion collisions.  The plotting style is the same as Fig.~\ref{fig:1}: the time slices are indicated by the colors, and the top and bottom panels correspond to the distributions in the $z$ and $y$ directions, respectively.  Note that the vertical scale is ten times greater in the bottom than in the top. 
}
\end{figure*}

The highly-charged matter created in intermediate-energy heavy-ion collisions (see Fig.~\ref{fig:1}) produces a strong electric field.  To show this, we plot, in Fig.~\ref{fig:2}, an electromagnetic Lorentz invariant, 
\begin{align}
	{\mathcal F} := {\bm E}^2 - {\bm B}^2 \;.
\end{align}
Note that the other invariant ${\mathcal G} := {\bm E} \cdot {\bm B}$ is found to be vanishing (up to residual event-by-event fluctuations).  This is reasonable, since we have no electric current ${\bm J}$ to produce a macroscopic magnetic field in central collision events and hence ${\bm B} \approx {\bm 0}$.  This also means
\begin{align}
	{\mathcal F} \approx {\bm E}^2\;.  
\end{align}
That is, the produced field is electric, which is also evident in Fig.~\ref{fig:2} as the sign of ${\mathcal F}$ is always positive.  It is thus convenient to introduce an {\it effective} electric field strength $E_{\rm eff}$ such that
\begin{align}
	E_{\rm eff} := \sqrt{{\mathcal F}} \;.
\end{align}

Figure~\ref{fig:2} shows that the typical strength of the produced electric field is $e^2{\mathcal F} = {\mathcal O}(10^{6\;\mathchar`-\; 7} \; {\rm MeV}^4)$, corresponding to $eE_{\rm eff} = {\mathcal O}((30\;\mathchar`-\; 60 \; {\rm MeV})^2)$.  The created field is strong in the sense that it is far beyond the Schwinger limit of QED, $eE_{\rm cr} = m_e^2 = (0.511\;{\rm MeV})^2$.  It is not supercritical to the hadron/QCD scale, $m_\pi \approx 140\;{\rm MeV}$, but still is non-negligibly strong $eE_{\rm eff} = {\mathcal O}((30\;\% \times m_\pi)^2)$, implying that it can affect hadron/QCD processes.  The produced field also has a sufficiently large spacetime volume $\tau \times V = {\mathcal O}(10\;{\rm fm}/c) \times {\mathcal O}((10\;{\rm fm})^3)$ because of the baryon stopping and of that the highly-charged matter has a macroscopically large spatial volume $\propto R^3/\gamma$.

Let us have a closer look at the spacetime structure of the produced field.  First, the produced field has a donut-like spatial structure, with peaks located around where the charge density vanishes (see Fig.~\ref{fig:1}).  Namely, the effective field strength $E_{\rm eff}$ is zero at the collision point ${\bm x} \approx {\bm 0}$, and, with increasing distance from the collision point, $E_{\rm eff}$ increases almost linearly $\propto |{\bm x}|$ and then decreases quadratically $\propto 1/|{\bm x}|^2$ after it peaks.  These are the direct consequences of the Gauss law ${\rm div}\,{\bm E} = \rho$ (and are similar to the textbook exercise of the electric field produced by a uniformly charged sphere; see, e.g., Jackson~\cite{Jackson:1998nia}).  

Second, the field strength in the transverse plane (the bottom panel in Fig.~\ref{fig:2}) increases with collision energy $\sqrt{s_{\rm NN}}$, while that along the beam direction (the top panel in Fig.~\ref{fig:2}) is less sensitive to $\sqrt{s_{\rm NN}}$ and is weaker than that in the transverse plane.  This is reminiscent of the anisotropic structure of the charged matter due to the Lorentz contraction (see Fig.~\ref{fig:1}) and can be understood simply with the Gauss law.  For simplicity, let us model the matter as a uniformly charged short cylinder, with radius $|{\bm x}_\perp| \leq R$ and length $|z| \leq R/\gamma$.  Then, the Gauss law tells us that the field strength is maximized along the longitudinal direction at the ends of the charged cylinder $|z| = R/\gamma$ and the corresponding strength is $E = \rho R/\gamma$.  Similarly, the maximum in the transverse plane is $E = \rho R/2$ at $|{\bm x}_\perp|=R$.  Thus, the field strength in the transverse plane is enhanced by the anisotropy factor $\gamma$ compared to that along the longitudinal axis.  

Note that the charge density $\rho$ can roughly be estimated as a simple overlapping of two Lorentz-contracted ions $\rho/e \approx \gamma\rho_0$ (see Sec.~\ref{sec:3A}).  Therefore, the maximum field strength over the space can be estimated, by using the simple modeling in the last paragraph, as $eE \approx e^2\gamma\rho_0R/2 \approx ((\sqrt{s_{\rm NN}}/1\;{\rm GeV})^{1/2} \times 31\;{\rm MeV})^2$.  This simple estimate is in rough agreement with the actual numbers shown in Fig.~\ref{fig:2} (albeit a bit overestimating, which we discuss in Sec.~\ref{sec:3C}).    

Third, the produced field expands faster with increasing collision energy.  For lower energies $\sqrt{s_{\rm NN}} \lesssim 5\;{\rm GeV}$, the expansion is negligible and the field is roughly staying around the original position created.  In contrast, for higher energies $\sqrt{s_{\rm NN}} \gtrsim 5\;{\rm GeV}$, the expansion becomes significant and the field moves almost at the speed of light.  For the longitudinal expansion, this is a natural consequence of the Bjorken picture beginning to be effective for higher energies.  Although the Landau picture dominates in the intermediate-energy regime as we have explained in Sec.~\ref{sec:3A}, it is also true that the Bjorken picture is becoming effective for higher energies, which can be observed in Fig.~\ref{fig:1} from that the charged matter expands in the beam direction after being created and begins to exhibit a two-peak structure at, e.g., $\sqrt{s_{\rm NN}} = 7.2\;{\rm GeV}$.  The expansion also becomes faster in the transverse direction, simply because the typical kinetic energy of the particles in the collision system increases with collision energy, allowing them to move faster.

\subsection{Field strength} \label{sec:3C}

\begin{figure*}[!t]
\flushleft{\hspace*{11mm} (i) Time evolution \hspace*{31mm} (ii) Peak strength \hspace*{41mm} (iii) Lifetime } \\
\vspace*{1mm}
\hspace*{-73mm}
\includegraphics[align=t, height=0.324\textwidth, clip]{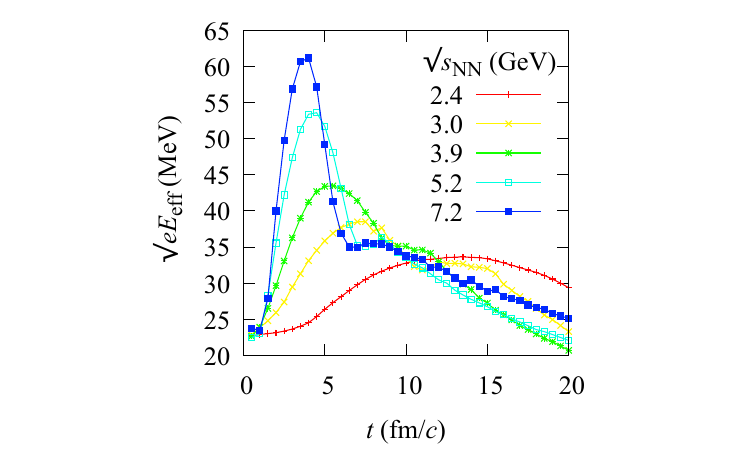} \hspace*{-7mm}
\includegraphics[align=t, height=0.31\textwidth, clip]{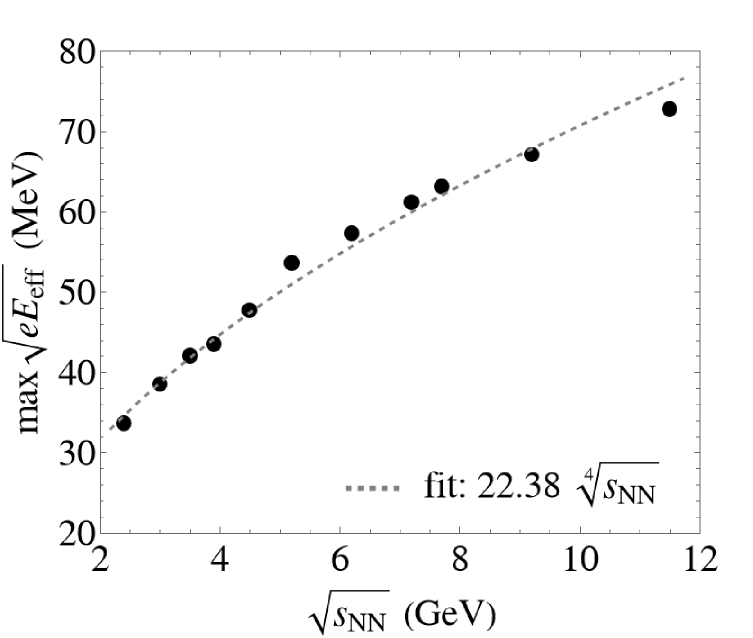} \hspace*{-4mm}
\includegraphics[align=t, height=0.31\textwidth, clip]{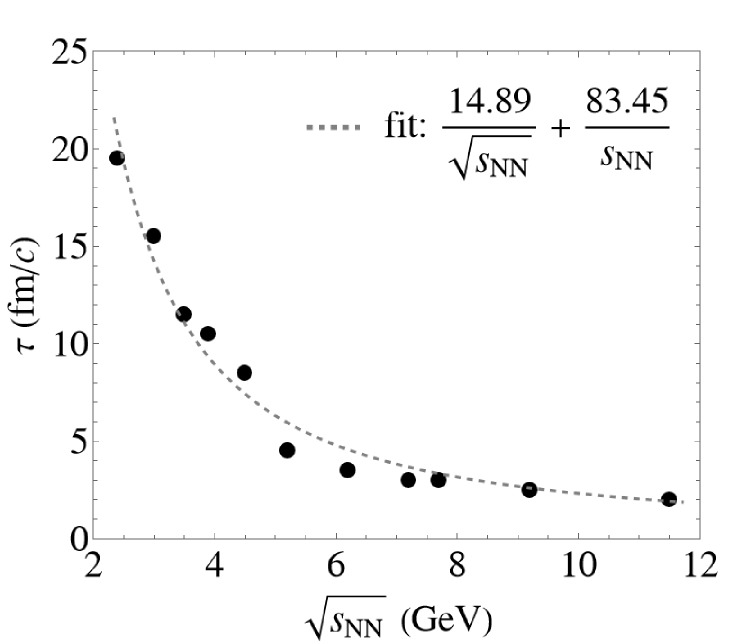}
\caption{\label{fig:3} The effective electric-field strength $E_{\rm eff}$ in intermediate-energy heavy-ion collisions.  (i) The time evolution of $E_{\rm eff}$ for some selected collision energies.  (ii) The maximum peak strength of $E_{\rm eff}$, extracted from (i), plotted against collision energy $\sqrt{s_{\rm NN}}$.  (iii) The lifetime of the field, defined as FWHM of $E_{\rm eff}$ (\ref{eq:10}).  The dashed lines in (ii) and (iii) are the numerical fittings of the obtained results.  }
\end{figure*}

Having observed the generation of a strong electric field in intermediate-energy heavy-ion collisions, we turn to discuss more quantitative aspects of the produced field, in particular the peak field strength over the space.  

We first discuss the time evolution of the field strength; see Fig.~\ref{fig:3} (i).  The strength gets maximized when the incident ions overlap with each other the most, and shows up as a sharp peak at around $t \approx (26\;{\rm fm}/c)/(\sqrt{s_{\rm NN}}/1\;{\rm GeV})$ (within the non-interacting estimate; see Sec.~\ref{sec:3A}).  After the peak, the field strength decays rather slowly, as the collision remnant can survive in the mid-rapidity region due to the baryon stopping, and consequently the field can maintain its strength for a relatively long time.  For example, a single ion at rest has a field strength of $eE \lesssim e^2 (\rho_0/2) R/3 \approx 25\;{\rm MeV}$, above which strength is kept for more than ${\mathcal O}(15\;{\rm fm}/c)$ in intermediate-energy heavy-ion collisions.  

The peak in Fig.~\ref{fig:3} (i) becomes larger and sharper with increasing collision energy.  This is also evident in Fig.~\ref{fig:3} (ii) and (iii), which show, respectively, the maximum strength of the peak, $\max E_{\rm eff}$, and the full width at the half maximum of $E_{\rm eff}$ (FWHM), 
\begin{align}
	\tau := \int_{E_{\rm eff}(t) > \max E_{\rm eff}/2 } {\rm d}t \;.  \label{eq:10}
\end{align}
Note that the FWHM (\ref{eq:10}) can naturally be regarded as the lifetime of the field produced, and so we use Eq.~(\ref{eq:10}) to {\it define} the lifetime in what follows.

The maximum peak field strength increases with collision energy [see Fig.~\ref{fig:3} (ii)], as anticipated with the simple estimate $eE \approx e^2\gamma\rho_0R/2 \approx ((\sqrt{s_{\rm NN}}/1\;{\rm GeV})^{1/2} \times 31\;{\rm MeV})^2$ that we made in Sec.~\ref{sec:3B}.  The key point is the enhancement of the charge density by the Lorentz contraction, which yields to the energy dependence.  Although the simple estimate can capture the qualitative feature of the peak field strength, it fails at the quantitative level mainly because the modeling with a uniformly-charged short cylinder is crude.  In particular, the actual charge distribution (see Fig.~\ref{fig:1}) is tailed and thus the charge extends more than the rigid cylinder distribution, which makes the actual peak field strength smaller than the simple estimate.  Indeed, we find
\begin{align}
	\max \sqrt{eE_{\rm eff}} = \left( \frac{\sqrt{s_{\rm NN}}}{1\;{\rm GeV}}\right)^{1/2} \times 22.38\;{\rm MeV} \;, \label{eq:11}
\end{align}
which has the smaller coefficient ($31\;{\rm MeV} \to 22\;{\rm MeV}$), can fit the numerical result well.

The lifetime (\ref{eq:10}) decreases with collision energy [see Fig.~\ref{fig:3} (ii)], which is essentially due to the Lorentz contraction.  The typical lifetime of the field is determined by that of the highly-charged matter.  Therefore, $\tau \approx 2R/\gamma \approx (26\;{\rm fm}/c)/(\sqrt{s_{\rm NN}}/1\;{\rm GeV})$ in the non-interacting limit (see Sec.~\ref{sec:3A}).  In reality, however, the interaction is important, especially for lower energies $\sqrt{s_{\rm NN}} \lesssim 5\;{\rm GeV}$, and thus the naive non-interacting estimate only gives a poor fit at the quantitative level.  The interaction makes the lifetime considerably longer.  We numerically find that such an interaction effect can be reproduced well with a higher-order term $\propto s_{\rm NN}^{-1}$, added to the naive $s_{\rm NN}^{-1/2}$ dependence, and the best fitting curve is found to be
\begin{align}
	\tau &= \left(\frac{\sqrt{s_{\rm NN}}}{1\;{\rm GeV}}\right)^{-1} \times 14.89\;{\rm fm}/c \nonumber\\
			&\quad + \left(\frac{\sqrt{s_{\rm NN}}}{1\;{\rm GeV}}\right)^{-2} \times 83.45\;{\rm fm}/c \;. \label{eq:12}
\end{align}
Note that the fitting curve (\ref{eq:12}) does not reproduce the coefficient of the naive estimate, $26\;{\rm fm}/c$, even in the limit of $\sqrt{s_{\rm NN}} \to \infty$.  This can be understood as an indication that the Bjorken picture cannot be complete and the interaction is important at intermediate energies.  

We emphasize that the lifetime is affected significantly by the interaction and made considerably longer, while the peak strength is less affected.  Accordingly, the lifetime is more dependent on collision energy than the peak strength is.  As we see shortly below, this shall be the essence why we can have a nonperturbatively strong electric field in intermediate-energy heavy-ion collisions, in particular for $\sqrt{s_{\rm NN}} \lesssim 5\;{\rm fm}/c$, where the interaction becomes more important.

\subsection{Nonperturbativity} \label{sec:3D}

\begin{figure}[!t]
\includegraphics[hsmash=c, width=0.5\textwidth]{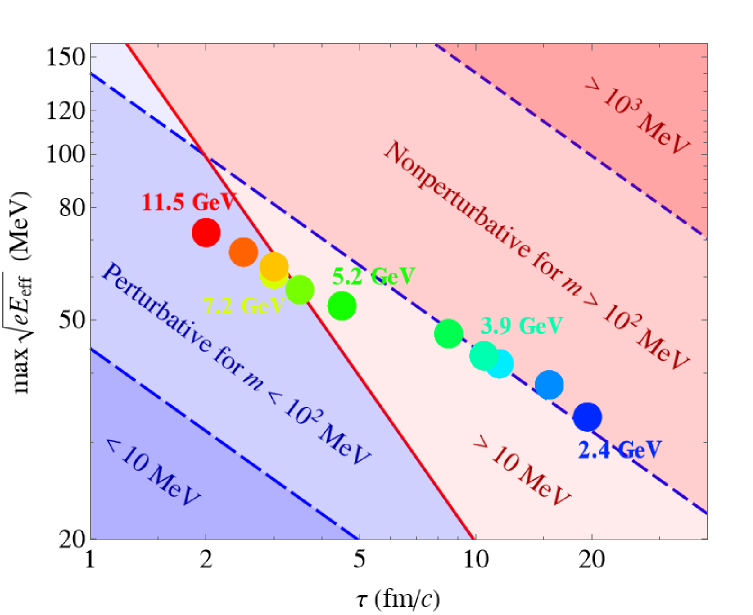} 
\caption{\label{fig:4} Sensitivity plot for nonperturbativity of the produced electric field in intermediate-energy heavy-ion collisions.  The dots represent the characteristics of the field $(\tau, \max \sqrt{eE_{\rm eff}})$ extracted from Fig.~\ref{fig:3} at each collision energy $\sqrt{s_{\rm NN}}$, ranging from $\sqrt{s_{\rm NN}} = 2.4\;{\rm GeV}$ (blue) to 3.0, 3.5, 3.9, 4.5, 5.2, 6.2, 7.2, 7.7, 9.2, and $11.5\;{\rm GeV}$ (red).  The lines represent the nonperturbativity parameters (\ref{eq:01}): $1 = \xi(10\;{\rm MeV})$ (bottom blue dashed), $1 = \xi(10^2\;{\rm MeV})$ (middle blue dashed), $1=\xi(10^3\;{\rm MeV})$ (top blue dashed), and $1 = \nu$ (red).  Those lines set ``phase boundaries'' of the nonperturbativity (of the vacuum pair production).  The red regions $\xi(m), \nu > 1$ are nonperturbative (for mass scales $m$), while the blue regions $\xi(m), \nu < 1$ are perturbative.  }
\end{figure}

As one of the possible ``order parameters'' for the nonperturbativity of a strong field, we calculate the nonperturbativity parameters $\xi$ and $\nu$ (\ref{eq:01}) and discuss the sensitivity region in intermediate-energy heavy-ion collisions.  The result is shown in Fig.~\ref{fig:4}.  It clearly shows that intermediate-energy heavy-ion collisions, in particular those with $\sqrt{s_{\rm NN}} \lesssim 5\;{\rm fm}/c$, can access the nonperturbative regime.  Namely, although higher collision energies $\sqrt{s_{\rm NN}} \gtrsim 5\;{\rm fm}/c$ are advantageous in that the field strength is strong, which can be comparable even to the hadron/QCD scale, it is disadvantageous in that the lifetime gets extremely short and accordingly the physics has to be purely perturbative.  On the other hand, although the achievable field strength is weaker for lower collision energies $\sqrt{s_{\rm NN}} \lesssim 5\;{\rm fm}/c$, it is still much stronger than the critical strength of QED and is non-negligible compared to the hadron/QCD scale, and also the lifetime can be very long, allowing us to enter the nonperturbative regime.  

Using the fitting results, Eq.~(\ref{eq:11}) for $\max \sqrt{eE_{\rm eff}}$ and Eq.~(\ref{eq:12}) for $\tau$, we find that the nonperturbativity parameters $\xi$ and $\nu$ can be parametrized as
\begin{align}
	\xi &= \frac{37.27\;{\rm MeV} + \left( \frac{\sqrt{s_{\rm NN}}}{1\;{\rm GeV}} \right)^{-1} \times 208.93\;{\rm MeV}}{m} \;, \\
	\nu &= \left( 9.34 \times s_{\rm NN}^{-3/4} + 1.67 \times s_{\rm NN}^{-1/4}  \right)^2 \approx 87.18 \times s_{\rm NN}^{-3/2} \;. \nonumber
\end{align}
It is clear that both $\xi$ and $\nu$ increase with decreasing collision energy $\sqrt{s_{\rm NN}}$, i.e., lowering the energy is more beneficial for the nonperturbativity.  The interaction effect in the Landau stopping regime gives a significant contribution here, as the dominant $s_{\rm NN}^{-1/2}$ and $s_{\rm NN}^{-3/2}$ dependencies in $\xi$ and $\nu$, respectively, arise from the $s_{\rm NN}^{-1}$ term in the lifetime $\tau$ (\ref{eq:12}), which we have added to account for the interaction effect.  

For clarity, we remark that the meaning of the ``nonperturbative'' in Fig.~\ref{fig:4} is just that the physical observable (or, to be precise, the number of particle and anti-particle pairs with mass $m$ produced from the vacuum by a strong electric field $E$~\cite{Popov:1971, Popov:1971iga, 1970PhRvD...2.1191B, Dunne:2005sx, Dunne:2006st, Oka:2011ct, Taya:2014taa, Gelis:2015kya, Aleksandrov:2018zso, Taya:2020dco}) acquires a nonperturbative dependence $\propto \exp[ -({\rm const})\times m^2/eE ]$.  In other words, being in the nonperturbative regime does not necessarily guarantee the significance of the nonperturbative effect and/or its detectability in actual experiments.  In fact, unless we have a field strength comparable to or exceeding the mass scale $eE \gtrsim m^2$, the effect is strongly suppressed by the exponential $\propto \exp[ -({\rm const})\times m^2/eE ]$ and is simply negligible.  For example, the lowest energy $\sqrt{s_{\rm NN}} = 2.4\;{\rm GeV}$ is nonperturbative for $m \approx 100\;{\rm MeV}$, but the field strength is at most $eE = {\mathcal O}((30\;{\rm MeV})^2)$ and therefore the corresponding nonperturbative effect is suppressed as $\propto \exp[ - (100/30)^2 ] = {\mathcal O}(10^{-5})$.  The exponential suppression becomes mild for $m^2 \lesssim eE$ and hence $m \lesssim 30\;{\rm MeV}$ is the mass region for $\sqrt{s_{\rm NN}} = 2.4\;{\rm GeV}$ such that the nonperturbative effect becomes significant and/or would be detectable in actual experiments.  In this sense of the significance/detectability, the highest collision energy closest to the phase boundary is the most advantageous to observe a nonperturbative effect with the largest $m$.  According to Fig.~\ref{fig:4}, such a collision energy is $\sqrt{s_{\rm NN}} \approx 5\;{\rm GeV}$ and the corresponding mass region is $m \lesssim 50\;{\rm MeV}$.  

Note also that the parameters $\xi$ and $\nu$ (\ref{eq:01}) are, to be precise, the nonperturbativity ``order parameters" for {\it temporal} inhomogeneity.  {\it Spatial} inhomogeneity also affects the nonperturbativity in general.  It is, however, less understood and the nonperturbativity for space- and spacetime-dependent fields is actually an open issue in strong-field QED; see Refs.~\cite{Dunne:2005sx, Dunne:2006st, Ilderton:2014mla, Gies:2015hia, Gies:2016coz, Fedotov:2022ely} for recent attempts.  To make a clean discussion, we therefore focused on the temporal inhomogeneity and used the parameters $\xi$ and $\nu$ in the present paper; it is our next step to estimate the spatial effects.  Nonetheless, we note that the spatial size of the produced field in intermediate-energy heavy-ion collisions ${\mathcal O}(10\;{\rm fm})$ (see Fig.~\ref{fig:2}) is rather large, compared to the lifetime, and therefore we would expect that the naive use of Eq.~(\ref{eq:01}) suffices at least for the purpose of judging the nonperturbativity.  Indeed, although the spatial size ${\mathcal O}(10\;{\rm fm})$ can be smaller than the Compton wavelength of low-mass particles [e.g., ${\mathcal O}(100\;{\rm fm})$ for a low momentum electron], it does not mean that the field is so small that it cannot nonperturbatively modify the particles' dynamics.  What matters is the work done by the field, which is not determined solely by the spatial extent but by the product between the field strength and the distance over which the electromagnetic force is exerted, and the particles' dynamics is modified drastically once the work exceeds the mass, i.e., $eE \lambda \gtrsim m$, with $\lambda$ being the spatial size.  The produced field strength is of the order of $eE_{\rm eff} = {\mathcal O}((30\;\mathchar`-\; 60 \; {\rm MeV})^2)$ (see Figs.~\ref{fig:2} and \ref{fig:3}), and therefore the work is ${\mathcal O}(50\;\mathchar`-\; 200 \; {\rm MeV})$.  This is large compared to electron- and even to pion-mass scales.

\section{Summary and discussion} \label{sec:4}
We have studied the generation of a strong electric field in head-on collisions of gold ions at intermediate energies ${\mathcal O}(3 \;\mathchar`-\; 10\;{\rm GeV})$, based on a hadron transport-model simulation JAM.  Our main statement is that intermediate-energy heavy-ion collisions are useful not only to study the densest matter on Earth but also to study strong-field physics in the nonperturbative regime.  Namely, we have shown that the produced electric field is as strong as $eE_{\rm eff} = {\mathcal O}((30\;\mathchar`-\; 60 \; {\rm MeV})^2)$ (see Figs.~\ref{fig:2} and \ref{fig:3}), which is supercritical to the Schwinger limit of QED and is still non-negligibly strong compared to the hadron/QCD scale.  The produced field is sufficiently long-lived and enables us to explore the nonperturbative regime of strong-field physics up to mass scale of $m\lesssim 50\;{\rm MeV}$ (see Fig.~\ref{fig:4}), which is inaccessible with any other experiments at present, such as high-power lasers and high-energy heavy-ion collisions.  We stress that this work is not just a re-estimation of the electromagnetic field in heavy-ion collisions with a never-used model.  As addressed in Sec.~\ref{sec:1}, the intermediate energy regime $\sqrt{s_{\rm NN}} = {\mathcal O}(3\;\mathchar`-\;10 \; {\rm GeV})$ (at central events) has never been studied in a quantitative manner in the literature, unlike the low- and high-energy regimes.  Our energy of interest is not a trivial issue but has physical significance, as the produced electric field produced has been shown to have very different characteristics compared to the other energy regimes and this is the reason why it is intriguing for both strong-field and hadron/QCD physics.  

The present work is just a first step, e.g., towards realistic estimations of nonperturbative strong-field effects and/or modifications to hadron/QCD dynamics in intermediate-energy heavy-ion collisions.  Let us briefly discuss possible directions and outlook below (in addition to those we have already mentioned; e.g., Sec.~\ref{sec:2b} on the parameter setting and Sec.~\ref{sec:3D} for the nonperturbativity due to spatial inhomogeneity).  

First, it is a very exciting possibility that we can study strong-field physics by using intermediate-energy heavy-ion collisions, since such a truly strong-field regime above the QED critical field strength cannot be achieved with any other experiments at the present.  One of the most intriguing targets is the Schwinger effect, i.e., the nonperturbative vacuum pair production, which was predicted more than seventy years ago~\cite{Sauter:1931zz, Schwinger:1951nm} and applied to various contexts, including the Hawking radiation in a black hole~\cite{Hawking:1975vcx}, but has never been observed in actual experiments yet.  As we have shown, intermediate-energy heavy-ion collisions generate a nonperturbatively strong electric field that is able to induce the Schwinger effect, and therefore can in principle be used as a new experimental setup to test it.  It is, therefore, an important task to predict possible experimental signatures of the Schwinger effect in heavy-ion collisions.  Naively, we can expect a thermal-like excess in electron/positron yields (and possibly muon/anti-muon as well) in the very low-momentum regime $|{\bm p}| \lesssim \sqrt{eE}$, where $eE \approx {\mathcal O}((50\;{\rm MeV})^2)$ for intermediate energies, as the celebrated Schwinger formula~\cite{Schwinger:1951nm} says that the momentum spectrum of the produced pairs is $\propto \exp[ -\pi (m^2+{\bm p}^2)/eE ]$.  It is worthwhile to quantify this excess, the zero-th order evaluation of which can be done with the Schwinger formula, or the locally-constant-field approximation~\cite{Bulanov:2004de, Gies:2016yaa, Gavrilov:2016tuq, Karbstein:2017pbf, Aleksandrov:2018zso, Sevostyanov:2020dhs}.  For a more realistic estimation, it would be necessary to include various modifications\footnote{The backreaction effect~\cite{Kluger:1991ib, Kluger:1992md, Kluger:1992md, Bloch:1999eu, Tanji:2008ku} may well be neglected in our situation due to the smallness of the coupling constant $\alpha := e^2/4\pi = 1/137$ for QED and $\alpha \approx 0.4$ for QCD at the energy scale $Q^2 \approx 1\;{\rm GeV}^2$~\cite{Deur:2016tte}.  It has been estimated in Ref.~\cite{Tanji:2008ku} that the typical time scale for when the backreaction effect becomes significant is $t \approx \sqrt{2\pi^2 /\alpha} (eE)^{-1/2} {\rm e}^{\pi m^2/2eE}$, which corresponds to $\approx 50 \times (eE)^{-1/2} \approx 50 \times (eE)^{-1/2} \approx 1500\;{\rm fm}$ for QED and $t \approx 200\;{\rm fm}$ for QCD (in both $m^2 \ll eE$ is assumed).  Those time scales are much longer than the lifetime of the electromagnetic field and also than the typical time scale for the heavy-ion reaction.  Therefore, the backreaction of the Schwinger effect is unimportant in heavy-ion collisions.  } due to, e.g., spatial inhomogeneity, polarization of the field, and event-by-event fluctuations in actual experiments.  For example, spatial inhomogeneity and/or polarization would modify not only the momentum spectrum but also spin/chirality/helicity imbalance~\cite{Strobel:2014tha, Wollert:2015kra, Ebihara:2015aca, Huang:2019szw, Huang:2019uhf, Taya:2020bcd, Takayoshi:2020afs, Aleksandrov:2024cqh}, which may be of interest from the viewpoint of the recent spin-polarization measurement (see Ref.~\cite{Huang:2020dtn, Becattini:2020ngo, Niida:2024ntm} for review and also Ref.~\cite{STAR:2021beb} for the investigation in the intermediate-energy regime).  Meanwhile, from an experimental point of view, low-momentum spectrum would naturally be contaminated by hadronic processes such as the Dalitz decay, $\pi^0 \to \gamma + e^- + e^+$, which poses a challenge how to eliminate those backgrounds to test the Schwinger effect with heavy-ion collisions.  

Second, from the standpoint of studying the densest matter and/or the QCD phase diagram with intermediate-energy heavy-ion collisions, the existence of the strong electric field is a noise that needs to be tamed, since electromagnetic observables such as charged flow and di-lepton yields are naturally affected.  For example, it has been argued in Ref.~\cite{Nishimura:2023oqn} that di-lepton yields are considerably enhanced due to the existence of the QCD critical point in the low invariant-mass regime $\lesssim 100\;{\rm MeV}$ and that such an enhancement can be used as an experimental signature of the QCD critical point.  Meanwhile, it is natural to expect additional contributions to such a low-energy di-lepton spectrum due to strong-field effects, including the Schwinger effect that we have mentioned in the last paragraph and other effects such as the nonlinear Breit-Wheeler process $\gamma + E \to e^+ + e^-$, but they have never been estimated.  Considering the fact that so far no clear signals of the critical point, or novel phases of QCD, have been found in the Beam Energy Scan program at RHIC~\cite{Aparin:2023fml}, it is reasonable to assume that the signals, if they exist, should be small.  Therefore, possible contaminations must be removed beforehand as much as possible, which of course applies to the strong-field effects.  

Third, our result suggests that intermediate-energy heavy-ion collisions may provide a unique opportunity to study QCD in a new extreme condition characterized by a strong electric field.  Although it would be difficult to have something very nontrivial for the hadronic scale $m = {\mathcal O}(100 \;\mathchar`-\; 1000\;{\rm MeV})$, which is greater than the achievable field strength $eE = {\mathcal O}((30\;\mathchar`-\; 60 \; {\rm MeV})^2)$, it is very reasonable to expect nonperturbative changes in the deconfined phase of QCD, where the typical mass scale $m$ is the current quark mass ${\mathcal O}(1\;{\rm MeV})$.  There are a number of studies of QCD in a strong magnetic field and it has been predicted that the QCD phase diagram is modified significantly (see, e.g., Ref.~\cite{Andersen:2014xxa}).  In contrast, only a few exist for a strong electric field~\cite{Yamamoto:2012bd, Endrodi:2023wwf} and there has been no consensus on what would happen, meaning that it requires further theoretical study.  

Fourth, our JAM simulation is in the default setting (see Sec.~\ref{sec:2b}) and therefore the presented results should be regarded as a {\it baseline} (based on the basic hadronic cascade model, widely used in the heavy-ion community) before the inclusion of nontrivial physical effects such as equation of state and hydrodynamic collective effects.  It is natural to expect that such effects become relevant at the high-density limit, and therefore it is important to include them for a more realistic estimation of the electromagnetic field.  At the moment, there is huge uncertainty in theory and phenomenological modeling of the dense hadron/QCD matter~\cite{Nagata:2021ugx}.  Thus, there do not exist any complete transport-model simulations for intermediate-energy heavy-ion collisions, which means that we need to wait for such development to improve our estimation.  One possible modification from such a beyond-naive-cascade modeling, the conductivity of the dense QCD/hadron matter may change the lifetime of the produced field~\cite{Tuchin:2010vs, McLerran:2013hla, Gursoy:2014aka, Tuchin:2015oka, Li:2016tel, Stewart:2021mjz}.  Another possibility is the impact of collective flow (e.g., modification to the directed flow $v_1$ due to the softening of the equation of state~\cite{Nara:2016phs}), which would enhance/reduce the anisotropy of the produced field and also the field strength due to the change of the Lorentz-contraction effect.  That the electromagnetic field is modified means that the electromagnetic probes should be modified also.  Conversely, it would be used to diagnose the properties of the dense QCD/hadron matter via the electromagnetic probes by comparing with the change from the naive baseline estimation, which is interesting as a novel way of studying the dense QCD/hadron matter.

\section*{Acknowledgments}
The authors thank Kensuke~Homma, Asanosuke~Jinno, Masakiyo~Kiatazawa, and Yasushi~Nara for enlightening discussions.  This work is supported by JSPS KAKENHI under grant No. 22K14045 (HT), the RIKEN special postdoctoral researcher program (HT), JST SPRING (grant No. JPMJSP2138) (TN), and the Multidisciplinary PhD Program for Pioneering Quantum Beam Application (TN).  

\bibliography{bib}
\end{document}